\documentstyle[12pt,epsfig]{article}
\begin{document}
\title{Quantum nondemolition measurements and non--Newtonian gravity.}
\author{ A. Camacho
\thanks{email: acamacho@nuclear.inin.mx} \\
Instituto Nacional de Investigaciones Nucleares. \\
Apartado Postal 18--1027, M\'exico, D. F., M\'exico.}

\date{}                                
\maketitle

\begin{abstract} 
In the present work the detection, by means of a nondemolition measurement, of a
Yukawa term, coexisting simultaneously with gravity, has been considered. In other words, a nondemolition
variable for the case of a particle immersed in a gra\-vi\-tational
field containing a Yukawa term is obtained. Afterwards the conti\-nuous monitoring
of this nondemolition parameter is analyzed, the co\-rres\-ponding propagator
is evaluated, and the probabilities associated with the po\-ssi\-ble measurement outputs
are found. The relevance of these kind of proposals in connection with some unified
theories of elementary particles has also been underlined.
\end{abstract}
\bigskip
\bigskip

\section{Introduction}
\bigskip

The equivalence principle (EP) is one of the fundamental cornerstones in modern physics,
and comprises the underlying symmetry of general relativity (GR) [1].
 At this point we must be more precise and state that EP has three different formulations,
 namely the weak, the medium strong, and finally, the very strong equivalence prin\-ci\-ple.
In order to avoid misunderstandings, here we follow [1], namely weak equivalence principle (WEP) means {\it the motion of any freely
falling test particle is independent of its composition and structure},
medium strong form (MSEP) means
{\it for every pointlike event of spacetime, there exists a sufficiently small
neighborhood such that in every local, freely falling frame in that neighborhood,
all the nongravitational laws of physics obey the laws of special re\-la\-ti\-vity}.
Replacing {\it all the nongravitational laws of physics} with {\it all the laws
of physics} we have the very strong form of the equivalence principle (VSEP).

The proposals
that confront the predictions of GR with measurement outputs include already a large amount
of experiments, for instance, the gravitational time dilation measurement [2],
the gravitational deflection of electromagnetic waves
[3], the time delay of electromagnetic waves in the field of the sun [4], or the
geodetic effect [5].
The discovery of the first binary pulsar PSR1913+16 [6]
allowed not only to probe the propagation properties of the gravitational
field [7], but it also offered the possibility of testing the case of strong
field gravity [8].
Of course, all these impressive experiments are an indirect confirmation of
the different EP.

Another important experimental direction comprises the attempts to test,
directly, WEP. Though these efforts are already more than a century old [9], the interest in this area has
not disappeared. Recently [10], WEP has been tested using a rotating 3 ton $^{238}$U
attractor around a compact balance containing Cu and Pb test bodies. The differential acceleration
of these test bodies toward the attractor was measured, and compared with the corresponding
gravitational acceleration. Clearly, this proposal is designed to test WEP at classical level,
i.e., gravity acts upon a classical system. At quantum realm the gravitational
acceleration has been measured using light pulse interferometers [11], and also
by atom interferometry based on a fountain of laser--cooled atoms [12]. Of course,
the classical experiment by Colella, Overhauser, and Werner (COW) [13], is also an experiment
that explores the effects at quantum level of gravity, and shows that at this level the
effects of gravity are not purely geometric [14].

The interest behind these experiments stems from the fact that various theoretical
attempts to construct a unified theory of elementary particles predict the existence
of new forces, and they are usually not described by an inverse--square law,
and of course, they violate one of the formulations of EP. By studying these violations one could determine
what interaction was producing these effects [15].

Among the models that in the direction of noninverse--square forces currently exist we have Fujii's proposal [16], in which a ``fifth force'', 
coexisting simultaneously with gravity, comprises a Yukawa term, 
$V(r) = -G_{\infty}{mM\over r}\Bigl(1 + \alpha e^{-{r\over \lambda}}\Bigr)$, 
here $G_{\infty}$ describes the interaction between $m$ and $M$ in the 
limit case $r\rightarrow\infty$, i.e., $G = G_{\infty}(1 + \alpha)$, 
where $G$ is the Newtonian gravitational constant. This kind of deviation terms arise from  
the exchange of a single new quantum of mass $m_5$, where the Compton wavelength 
of the exchanged field is $\lambda = {\hbar\over m_5c}$ [15], this field is usually denoted dilaton. 

The experiments, already carried out, that intend to detect a Yukawa term have already imposed
some limits on the parameters $\alpha$ and $\lambda$. For instance,
if $10^{-4}m \leq \lambda \leq 10^{-3}m$, then $\alpha \sim 10^{22}$ [17],
if $\lambda = 200\mu M$, then $\alpha \leq 8\times 10^{7}$ [18] (for a more
complete report see [15]).

To date, after more than a decade of experiments [19], there is no compelling e\-vi\-den\-ce 
for any kind of deviations from the predictions of Newtonian gravity. 
But Gibbons and Whiting  (GW) phenomenological analysis of gravity data [20] has 
proved that the very precise agreement between the predictions of Newtonian gravity 
and observation for planetary motion does not preclude the existence of large 
non--Newtonian effects over smaller distance scales, i.e., precise experiments 
over one scale do not ne\-ce\-ssarily constrain gravity over another scale. 
GW results conclude that the current ex\-pe\-ri\-men\-tal constraints over possible 
deviations did not severly test Newtonian gra\-vi\-ty over the $10$--$1000$m distance 
scale, usually denoted as the ``geophysical window''.

The idea in this work is two--fold: firstly, the effects of
a Yukawa term upon a quantum system (the one is continuously monitored)
will be calculated; secondly, new theoretical predictions for one
of the models in the context of quantum measurement theory will be found.
In order to achieve these two goals we will obtain a nondemolition variable for the case of a particle
subject to a gravitational field which contains a Yukawa term such that
$\lambda$ has the same order of magnitude of the radius of the earth. At
this point it is noteworthy to mention that the current experiments set constraints
for $\lambda$ for ranges between 10km and 1000km [10], but
the case in which $\lambda \sim$ Earth's radius remains rather unexplored. Afterwards, we will
consider, along the ideas of the so--called restricted path integral formalism
(RPIF) [21], the continuous monitoring of this nondemolition parameter, and calculate, not only,
the corresponding propagators, but also the probabilities associated with the
diffe\-rent measurement outputs.
\bigskip

\section{Yukawa term}
\bigskip

Suppose that we have a spherical body with mass $M$ and radius $R$.
Let us now consider the case of a Yukawa form of gravitational interaction [16],
hence the gravitational potential of this body reads

{\setlength\arraycolsep{2pt}\begin{eqnarray}
V(r) = -G_{\infty}{M\over r}\Bigl(1 + \alpha e^{-{r\over \lambda}}\Bigr).
\end{eqnarray}

Let us now write $r = R + z$, where $R$ is the body's radius, and $z$ the height 
over its surface. If $R/\lambda \sim 1$ (which means that the range
of this Yukawa term has the same order of magnitude as the radius of our
spherical body), and if $z<<<R$, then we may approximate the Lagrangian of a particle of mass $m$ as follows

{\setlength\arraycolsep{2pt}\begin{eqnarray}
L = {{\vec{p}}~^2\over 2m} + G_{\infty}{mM\over R}
\left([1 + \alpha] - [{1 + \alpha\over R} + {\alpha\over 2\lambda}]z
+[{1 + \alpha\over 2R^2} + {\alpha\over 2R\lambda}+ {\alpha\over 2\lambda^2}]z^2\right).
\end{eqnarray}
\bigskip

\section{Quantum Measurements}
\bigskip

Nowadays one of the fundamental problems in modern physics comprises
the so--called quantum measurement problem [22]. Though there are several attempts to solve this
old conundrum (some of them are equivalent [23]), here we will resort to
RPIF [21], because it allows us to calculate, in an easier manner, propagators
and pro\-ba\-bi\-lities. RPIF explains a continuous quantum measurement with the introduction of a restriction on 
the integration domain of the corresponding path integral.
This last condition can also be reformulated in terms
of a weight functional that has to be considered in the path integral.
Clearly, this weight functional contains 
all the information about the interaction between measuring device and measured system.
This model has been employed in the analysis of the response of a gravitational
antenna not only of Weber type [21], but also when the measuring process involves
a laser--interferometer [24]. We may also find it in the quest for an explanation
of the emergence of some classical properties, as time, in quantum cosmology [25].

Suppose now that our particle with mass $m$ goes from point $N$ to point $W$.
Hence its propagator reads

{\setlength\arraycolsep{2pt}\begin{eqnarray}
U(W,\tau'';N, \tau') = \left({m\over 2\pi i\hbar T}\right)
\exp\left\{{im\over 2\hbar T}\left[(x_W - x_N)^2 + (y_W - y_N)^2\right]\right\}\nonumber\\
\times\int_{z_N}^{z_W}d[p]d[z(t)]\exp\left\{{i\over\hbar}\int_{\tau'}^{\tau''}
\left[{p^2\over 2m} + (1 + \alpha){G_{\infty}mM\over R} + Fz + {m\over 2}\Omega^2z^2\right]dt\right\}.
\end{eqnarray}}

Here we have introduced the following definitions

{\setlength\arraycolsep{2pt}\begin{eqnarray}
F = -G_{\infty}{mM\over R}\left[{1 + \alpha\over R} + {\alpha\over 2\lambda}\right],
\end{eqnarray}}

{\setlength\arraycolsep{2pt}\begin{eqnarray}
\omega^2 = -2{G_{\infty}M\over R}\left[{1+ \alpha\over 2R^2}+
{\alpha\over 2\lambda R} + {\alpha\over 2\lambda^2}\right].
\end{eqnarray}}

$T = \tau'' - \tau'$, and $\sqrt{(x_W - x_N)^2 + (y_W - y_N)^2}$ denotes the projection on the
body's surface of the distance between points $W$ and $N$. We also have that $G = G_{\infty}[1 + \alpha]$, and $G$ is the Newtonian gravitational constant [15].
In our case, $[{1+ \alpha\over 2R^2}+
{\alpha\over 2\lambda R} + {\alpha\over 2\lambda^2}] >0$, hence, $\omega = i\Omega$, where $\Omega\in\Re$.
\bigskip

Suppose now that the variable $A(t)$ is continuously monitored. Then we must consider,
along the ideas of RPIF, a particular expression for our weight functional, i.e., for $w_{[a(t)]}[A(t)]$. 
As was mentioned before, the weight functional $w_{[a(t)]}[A(t)]$ contains the information concerning the measuring process. 

At this point we face a problem, namely, the choice of our weight functional.
In order to solve this difficulty, let us mention that the results coming from a Heaveside weight functional [26] and those
coming from a gaussian one [27] coincide up to the order of magnitude. 
These last remarks allow us to consider a gaussian weight functional as an 
approximation of the correct expression.
But a sounder justification of this choice stems from the fact that there 
are measuring processes in which the weight functional possesses a gaussian form [28]. 
In consequence we could think about a measuring device whose weight functional is very close to a gaussian behaviour.

Therefore we may now choose as our weight functional the following expression 

\begin{equation}
\omega_{[a(t)]}[A(t)] = \exp\left\{-{2\over T\Delta a^2}\int _{\tau '}^{\tau ''}\left[A(t) - a(t)\right]^2dt\right\},
\end{equation}
\bigskip

\noindent here $\Delta a$ represents the error in our measurement, i.e., it is the resolution of the measuring apparatus.
\bigskip

\section{Quantum nondemolition measurements}
\bigskip

The basic idea around the concept of quantum nondemolition (QND) measurements
is to carry out a sequence of measurements of an observable in such a way that
the measuring process does not diminish the predictability of the results of subsequent measurements of the same observable [29].
This concept stems from the work in the context of gravitational wave antennae.
Indeed, the search for gravitational radiation demands measurements of very small
displacements of macroscopic bodies [30].  Braginsky {\it et al} [31] showed that there
is a quantum limit, the so called ``standard quantum limit'', which is a
consequence of Heisenberg uncertainty principle, the one limits the sensitivity of
the corresponding measurement (the original work [31] involves the sensitivity of
a gravitational antenna). This work allowed also the introduction of the idea of a QND measurement,
in which a variable is measured in such a way that the unavoidable disturbance
of the conjugate observable does not disturb the evolution of the chosen variable [32].

Let us now suppose that in our case $A(t) = \rho p + \sigma z$, where $\rho$ and $\sigma$ are functions of time.
In this particular case, the condition that determines when $A(t)$ is a QND
variable may be written as a differential equation [21]

\begin{equation}
{df\over dt} = {f^2\over m} - m\Omega^2,
\end{equation}
\bigskip

\noindent where $f(t)=  \sigma/\rho$. 
\bigskip

It is readily seen that a solution to (7) is

\begin{equation}
f(t)= -m\Omega\tanh(\Omega t).
\end{equation}
\bigskip

Choosing $\rho(t) = 1$, we find that in our case a possible QND variable is

\begin{equation}
A(t) = p - m\Omega z\tanh(\Omega t).
\end{equation}
\bigskip

\section{QND and non--Newtonian gravity: Propagators and probabilities}
\bigskip

With our weight functional choice (expression (6)) the new propagator involves two gaussian integrals, and can be easily calculated [33]

{\setlength\arraycolsep{2pt}\begin{eqnarray}
U_{[a(t)]}(W,\tau'';N, \tau') = \left({m\over 2\pi i\hbar T}\right)
\exp\left\{{im\over 2\hbar T}\left[(x_W - x_N)^2 + (y_W - y_N)^2\right]\right\}\nonumber\\
\exp\left\{{i\over\hbar}(1 + \alpha){G_{\infty}mM\over R}T\right\}
\exp\left\{{-T\Delta a^2 + i2m\hbar\over 4m^2\hbar^2 + T^2\Delta a^4}\int_{\tau'}^{\tau''}a^2(t)dt\right\}\nonumber\\
\times\exp\{{-i\hbar\over 2m\Omega^2}\int_{\tau'}^{\tau''}\left[{F\over\hbar} + {4m^2\hbar\Omega a\over 4m^2\hbar^2 + T^2\Delta a^4}\tanh(\Omega t)
+ i{2ma\Omega T\Delta a^2\over 4m^2\hbar^2 + T^2\Delta a^4}\right]^2\nonumber\\
\times\left[{4m^2\hbar^2[1 + \tanh^2(\Omega t)] +
T^2\Delta a^4 - i2m\hbar T\Delta a^2\tanh^2(\Omega t)\over 4m^2\hbar^2[1 + \tanh^2(\Omega t)]^2 +
T^2\Delta a^4 }\right]dt\}.
\end{eqnarray}}     
\bigskip

The probability, $P_{[a(t)]}$, of obtaining as measurement output $a(t)$ is given by expression
$P_{[a(t)]} = \vert U_{[a(t)]}\vert^2$ [21].  Hence, in this case

{\setlength\arraycolsep{2pt}\begin{eqnarray}
P_{[a(t)]}=
\exp\left\{{-2T\Delta a^2 \over 4m^2\hbar^2 + T^2\Delta a^4}\int_{\tau'}^{\tau''}a^2(t)dt\right\}\nonumber\\
\times\exp\left\{{\hbar\over m\Omega^2}\int_{\tau'}^{\tau''}\left[2I_1I_2I_3 + I_4(I^2_2 - I^2_1)\right]dt\right\}.
\end{eqnarray}}     
\bigskip

Here the following definitions have been introduced

{\setlength\arraycolsep{2pt}\begin{eqnarray}
I _1 = {F\over\hbar} + {4m^2\hbar\Omega a\over 4m^2\hbar^2 + T^2\Delta a^4}\tanh(\Omega t),
\end{eqnarray}}

{\setlength\arraycolsep{2pt}\begin{eqnarray}
I _2 = {2ma\Omega T\Delta a^2\over 4m^2\hbar^2 + T^2\Delta a^4}\tanh(\Omega t),
\end{eqnarray}}

{\setlength\arraycolsep{2pt}\begin{eqnarray}
I _3 = {4m^2\hbar^2[1 + \tanh^2(\Omega t)] + T^2\Delta a^4\over 4m^2\hbar^2[1 + \tanh^2(\Omega t)]^2 + T^2\Delta a^4},
\end{eqnarray}}

{\setlength\arraycolsep{2pt}\begin{eqnarray}
I _4 = {2m\hbar T\Delta a^2\tanh^2(\Omega t)\over 4m^2\hbar^2[1 + \tanh^2(\Omega t)]^2 + T^2\Delta a^4}.
\end{eqnarray}}
\bigskip
\bigskip

\section{Conclusions}
\bigskip

In this work we have considered a Yukawa term coexisting with the usual Newtonian
gravitational potential (expression (1)). Assuming that the range of this new interaction has the
same order of magnitude than the Earth's radius a QND variable was obtained for
a particle with mass $m$ (expression (9)), and its corresponding propagator was evaluated.
Afterwards, it was assumed that this QND variable was continuously monitored, and,
along the ideas of RPIF, the propagator and proba\-bi\-lity  associated with the
possible measurement outputs were also calculated, expressions (14) and (15), respectively.

Another interesting point around expressions (10) and (11) comprises the role that the mass
parameter plays in them. It is readily seen that if we consider two particles with different mass, say
$m$ and $\tilde {m}$, then they render different propagators and probabilities.
This last fact means that ``gravity'' is, in this situation, not purely geo\-metric.
This is no surprise, the presence of a new interaction, coexisting with the
usual Newtonian gravitational potential, could mean the breakdown of the geometrization of the
joint interaction (Newtonian contribution plus Yukawa term). The detection of this interaction
would mean the violation of WEP, but not necessarily of VSEP [15].
Nevertheless, the possibilities of using quantum measurement theory to analyze
the possible limits of VSEP at quantum level do not finish here, indeed, the possible incompatibility
between the different formulations of EP and measurement theory can also be studied along the ideas of this theory [42]. Of course, more
work is needed around the validity at quantum level of the different formulations
of EP. Indeed, as has already been proved [43], in the
presence of a gravitational field, the generalization to
the quantum level of even the simplest kinematical concepts, for instance the time of flight, has severe conceptual
difficulties.

This proposal would also render new theoretical predictions that could be confronted (in the future)
against the experiment, and therefore we would obtain a larger framework that could allow us
to test the validity of RPIF [44]. As was mentioned before, this comprises our second goal
in the present work.
\bigskip

\Large{\bf Acknowledgments}\normalsize
\bigskip

The author would like to thank A. A. Cuevas--Sosa for his help. It is also a pleasure to thank
R. Onofrio for bringing reference [29] to my attention. This work was partially
supported by CONACYT (M\'exico) Grant No. I35612--E.

\end{document}